\documentclass{article}
\usepackage{spconfa4,amsmath,graphicx,bm,acronym,booktabs,balance,url}
\usepackage[table,xcdraw]{xcolor}

\title{Real-time Speech Restoration using Data Prediction Mean Flows}
\name{Sebastian Braun}
\address{Microsoft Research, Redmond, WA, USA}
\acrodef{FM}{Flow Matching}
\acrodef{CFM}{Conditional Flow Matching}
\acrodef{OT}{Optimal Transport}
\acrodef{DP}{data prediction}
\acrodef{ODE}{Ordinary Differential Equation}
\acrodef{JVP}{Jacobian-vector Product}
\acrodef{IMF}{Improved Mean Flow}
\acrodef{MF}{Mean Flow}
\acrodef{STFT}{short-time Fourier transform}
\acrodef{RIR}{room impulse response}
\acrodef{SNR}{signal-to-noise ratio}
\acrodef{NFE}{number of function evaluations}
\acrodef{TCN}{temporal convolutional network}
\acrodef{MAC}{multiply-accumulate operations}
\acrodef{WER}{word error rate}
\acrodef{SOTA}{state-of-the-art}

\begin{document}
\ninept
\maketitle
\begin{abstract}
Generative models are capable to address difficult problems with non-unique solutions like bandwidth extension and gap filling, removing highly non-linear artifacts from codecs, clipping and distortion, as opposed to removing linear additive components like noise and reverb. While large offline processing models have shown impressive results, these tasks have not been solved with real-time capable models with low latency and compute. We propose a few-step flow matching model using Data Prediction Mean Flows in combination with suitable novel low-latency architecture to make flow matching models an attractive choice under theses constraints. Compared to state-of-the-art, our proposed mean flow model uses 120x less compute and introduces no algorithmic latency other than the STFT, while achieving similar audio quality.
\end{abstract}
\section{Introduction}
\label{sec:intro}
While speech enhancement typically only treats linear additive degradations like background noise and reverberation, in many scenarios, the audio signals are recorded in sub-optimal acoustic conditions: destructive interference like wind noise, cheap or broken microphone and recording chains, heavy destructive or non-linear processing, and content loss from audio compression or network transmission are not uncommon in real-world consumer audio pipelines, greatly affecting speech signal quality. Speech restoration aims to address and fix all of those non-linear and destructive degradations by using generative modeling. In the past few years, generative models have become powerful enough to achieve impressive results on this task \cite{Su2021,Serra2022,Richter2024}. 

Although diffusion and \ac{FM} models mostly outperform the quality of GANs, \ac{FM} models achieving impressive and robust results are large offline processing methods only. However, most audio applications run under heavy constraints on devices and also are real-time streaming applications, e.\,g.\ in telecommunications, hearing assistive and augmented reality devices, or any other live audio processing or interactive task. 
Further, many approaches operate in a learned latent (VAE) or lossy spectral (Mel) domain, requiring additional encoders/decoders, which can incur additional complexity, latency, and add potential errors or hallucinations, which complicates optimization and real-time deployment. 

Most \ac{SOTA} audio models using diffusion or \ac{FM} build on either the \emph{NCSN++} \cite{Song2021} or diffusion transformer \cite{Evans2025}, where the latter is unattractive for real-time applications due to being an offline encoder. NCSN++ is a convolutional U-net suitable to stream infinite audio sequences. However, due to its internal temporal feature down-sampling, it incurs a large algorithmic latency of over 600~ms, prohibiting its application in latency critical real-time scenarios like communication, where audio processing latency is typically expected below 30~ms.
DiffusionBuffer \cite{Lay2025} attempts to reduce latency and complexity by folding the diffusion steps over audio time. However, as at still multiple diffusion steps are required for satisfactory performance, the resulting system still accumulates a latency of 180-320~ms. In \cite{Hsieh2026} we modify NSCN++ to be causal and remove its latency by removing temporal down-sampling. While the modification worked despite some performance drop, this also highly increased the complexity by 3x due to larger internal feature maps. Therefore, we also explored a new much smaller U-net architecture for \ac{FM}, but it significantly under-performed for the task of general speech restoration. In parallel, Welker et.\ al.\ \cite{Welker2025} proposed a more sophisticated modification to NCSN++, similarly making it causal, extending receptive field, using \ac{FM} objective, and using a more complex inference solver. While their streaming model showed less degradation vs.\ the non-causal NCSN++ model, the complexity increased even further by over 5x.

Another issue of diffusion and \ac{FM} is the complexity scaling with the number of inference steps, often referred to \ac{NFE}. While the today's standard \emph{optimal transport} \ac{CFM} training target in theory provides a linear trajectory from prior to target distribution, which would be solvable in a single step, in practice estimation errors still cause high trajectory curvature. Among \emph{rectified flow} and \emph{consistency models},
\emph{Mean Flow} \cite{Geng2025} is a newer and promising technique to reduce the \ac{NFE} by training larger steps instead of infinitesimal small steps. 
Mean Flow has been adopted for text-to-audio generation \cite{Li2025} and target speaker extraction \cite{Shimizu2025} and speech enhancement \cite{Li2026}, but neither in a real-time focused low latency/low compute mindset and not for generative speech restoration.

In this work, we advance speech restoration \ac{FM} models for real-world real-time deployment: i) As a first application in audio, we combine \ac{IMF} training with data prediction, replacing the instantaneous velocity loss, to boost the quality and reduce inference steps. ii) We further propose careful choices of flow time distribution, \ac{IMF} scheduling, and a prior distribution by starting at the noisy mean and using a spectrally similar noise to audio for performance gains. iii) We develop a new model backbone, with 120x complexity reduction and no latency penalty, and push its performance towards the complex, non-causal NCSN++ backbone with all training modifications. The results are demonstrated using large-scale training to verify generalization, and evaluated on real recordings using subjective and representative objective metrics.
 

\section{Method}
\label{sec:pagestyle}
Our \ac{FM} framework operates in the complex compressed spectral domain, obtained via \ac{STFT} and applying magnitude compression by
\begin{equation}
    \label{eq:complex_compressed_signal}
    X^c(k,n) = |X(k,n)|^c \; \frac{X(k,n)}{|X(k,n)|}
\end{equation}
where $k,n$ are the frequency and time indices, and we set the magnitude compression to $c=0.3$. Within the \ac{STFT} operation, the spectrum is power normalized using the window energy, so the maximal spectral amplitudes for bound digital signal within time domain amplitudes $[-1,1]$ do not exceed spectral magnitude 1 as well. This holds before and after magnitude compression. After processing by \ac{FM} models, the enhanced signal is transformed back to time domain with corresponding de-compression and inverse normalization.

\subsection{Flow Matching}

\ac{CFM} \cite{Lipman2023}, a deterministic version of diffusion \cite{Ho2020}, is a method to gradually transform a prior distribution $\bm{x}_1 \sim p_\text{init}$ to a posterior data distribution $\bm{x}_0 \sim p_\text{data}$ over flow time $t$. In accordance with Mean Flow literature, we use the reverse flow time definition $t: 1 \rightarrow 0$, which coincides with diffusion literature.

For speech enhancement, restoration or bandwidth extension tasks, where the conditon, here the degraded signal $\bm{y}$, are partially highly correlated,
\ac{CFM} using optimal transport often uses an informed prior $\bm{\mu}_1$, rather than starting from a zero-mean Gaussian distribution for $p_\text{init}$. Specifically, we set
\begin{equation}
    p_\text{init} = \mathcal{N}(\bm{\mu}_1, \sigma_\text{max} \bm{I})
\end{equation}

The \ac{OT} plan linearly moves the variable $\bm{x}_t$ from a noisy initial distribution, centered around the degraded audio data $\bm{y}$ with noise variance $\sigma_\text{max}$, to the data $\bm{x}_0$ with possibly a small residual noise with variance $\sigma_\text{min}$:
\begin{eqnarray}
    \bm{\mu}_t =& (1-t) \bm{x}_0 + t \bm{y} \label{eq:prior_mu}\\
    \sigma_t =& (1-t) \sigma_\text{min} + t \sigma_\text{max} \label{eq:prior_sigma}\\
    \bm{x}_t =& \bm{\mu}_t + \sigma_t \bm{\epsilon} \label{eq:prior_x}
\end{eqnarray}
where $\bm{\epsilon} \sim \mathcal{N}(0, \bm{I})$ is the "diffusion" noise corrupting the intial data and being removed gradually over time $t$, and $\bm{y}$ is the degraded input audio data. 
The vanilla \ac{FM} objective trains a model with parameters $\theta$ to predict the instantaneous flow target
\begin{equation}
    \bm{v}_t = (\sigma_\text{max} - \sigma_\text{min}) \bm{\epsilon} - (\bm{x}_0 - \bm{y})
\end{equation}
which in the case of \ac{CFM}-\ac{OT} is independent of $t$. 
A neural flow estimator $\bm{u}_\theta(\bm{x}_t, \bm{y}, t)$, conditioned on the flow data variable, the degraded signal and flow time, is optimized via regression loss
\begin{eqnarray}
    \label{eq:fm_loss}
    \mathcal{E}\{ \Vert \bm{v}_t - \bm{u}_\theta(\bm{x}_t, \bm{y}, t) \Vert^2 \}
\end{eqnarray}

For inference, the data estimate is obtained via the \ac{ODE}. We use the simple Euler integration:
\begin{equation}
    \label{eq:ode}
    \bm{x}_{t-1} = \bm{x}_t - \bm{u}_\theta(\bm{x}_t, \bm{y}, t) \, dt
\end{equation}

\subsection{Data Prediction Loss}
\label{sec:data_prediction}
An alternative \ac{CFM} solution is to optimize the neural estimator to predict the data $\bm{x}_0$ itself, resulting in the \ac{DP} loss
\begin{eqnarray}
	\label{eq:data_prediction_loss}
    \mathcal{E}\{ \Vert \bm{x}_0 - \bm{\hat x}_\theta(\bm{x}_t, \bm{y}, t) \Vert^2 \}
\end{eqnarray}
For inference, the data estimate is converted to the instantaneous flow via
\begin{equation}
    \label{eq:data_prediction_conversion}
    \bm{\hat v}_t = (\bm{x}_t - \bm{\hat x}_\theta(\bm{x}_t, \bm{y}, t)) / t
\end{equation}
and is then plugged into the \ac{ODE} \eqref{eq:ode}.
Note that this does not equal simple signal error based optimization, since the network is conditioned on $\bm{x}_t$ and $t$, and the curvature of the taken flow trajectory in practice still leads the model to a better generative solution than a non-generative single-step prediction model.
We find \ac{DP}-\ac{FM} better performing and more stable than the traditional velocity-\ac{FM} objective \eqref{eq:fm_loss}, especially for smaller models.

\subsection{Mean Flows with data prediction}
Mean flows \cite{Geng2026,Geng2025} intends to reduce the required inference step size of \ac{FM}: instead of taking infinitesimally small steps, it trains arbitrary large steps within $t \in [0,1]$ by introducing a second time conditioning, the start time $r \leq t$. The mean flow model is trained with arbitrary large steps from $r \rightarrow t$ by integrating the instantaneous flow $\bm{u}(r,t) = \int_r^t \bm{v}(\bm{x}_t)$
The tractable solution is obtained by the mean flow identity, adding the derivative w.r.t. $t$ to the loss
\begin{equation}
    \bm{u}(r,t) = \bm{v}_t - (t-r) \frac{d}{dt} \bm{u}(\bm{x}_t)
\end{equation}
The derivative $\frac{d}{dt} \bm{u}$ can be obtained by the \ac{JVP} between the Jacobian $[\delta_x u, \delta_r u, \delta_t u]$ and the tangent $[v, 0, 1]$ and is implemented in PyTorch. Note that for $r=t$, Mean Flow is identical to vanilla \ac{FM}.
We directly resort to the \ac{IMF} solution \cite{Geng2025}, which stabilizes the training objective by removing the training target dependency on the trained network itself. By re-parameterizing the mean flow identity, \ac{IMF} defines a prediction function for our network:
\begin{equation}
	\label{eq:imf_prediction}
    \bm{V}_\theta = \bm{u}_\theta + (t-r) \operatorname{JVP}_\text{sg}
\end{equation}
where $\operatorname{JVP}_\text{sg}$ is the JVP with stop-gradient operator. \ac{IMF} is characterized as u-prediction with v-loss.
Note that the tangent of the \ac{JVP} depends on the instantaneous flow $v$, which is obtained by evaluating the current network $\theta$ at $r=t$.

Finally, to combine the benefits of data prediction (x-loss) for traditional \ac{FM} outlined in Sec.~\ref{sec:data_prediction}, we follow [pixelMeanFlow] by re-defining our neural estimator as $\bm{x}_\theta(\bm{x}_t, y, t, r)$ to predict the target data $\bm{x}_0$. The network output can be converted to the instantaneous velocity via
\begin{equation}
    \bm{u}_\theta = (\bm{x}_t - \bm{x}_\theta(\bm{x}_t, \bm{y}, t, r)) /t 
\end{equation}
and then plugged into \ac{IMF} loss \eqref{eq:imf_prediction} as before. For inference, we similarly resort to the conversion network output to instantaneous flow \eqref{eq:data_prediction_conversion}, and compute the \ac{ODE} \eqref{eq:ode}.

\subsection{Training scheduling}
\begin{figure}[tb]
    \centering
    \includegraphics[width=0.9\columnwidth, clip, trim=0 4 0 2]{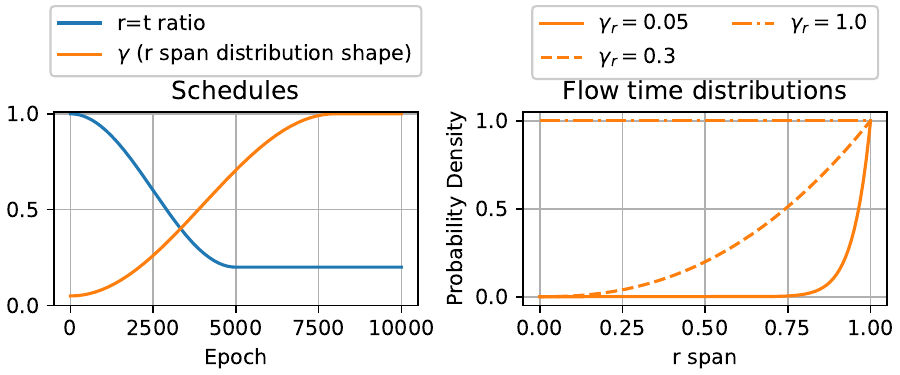}
    \caption{Training schedules for $r=t$ ratio and $r$-$t$-span sampling exponent $\gamma$}
    \label{fig:schedules}
\end{figure}

The JVP requires the derivative evaluated at $r=t$, so it is crucial to include this condition in the training process. In original \ac{MF} \cite{Geng2025} a collapse is reported if this condition is not large enough, so they set 75\% of training samples to $r=t$. The stabilization in \ac{IMF} reports a possible decrease of the $r=t$ percentage down to 25\%. We propose a sigmoid-based training schedule for the ratio as shown in Fig.~\ref{fig:schedules} blue.
MeanFlowSE \cite{Li2026} further proposes a scheduling of the span between $r$ and $t$. We start with a spiky distribution of smaller steps and increase the span gradually over training time by scheduling the time span distribution shape. The $r$-span factor $\Delta_r$ is sampled from a uniform distribution with exponent $0<\gamma \leq 1$ by
\begin{equation}
    \Delta_r \sim \mathcal{U}(0, t)^\gamma
\end{equation}
and we obtain a sampled $r \leq t$ for given sampled $t$ by
$r= \Delta_r \cdot t$.
We schedule the exponent $\gamma \in (0.05 \rightarrow 1)$ as shown Fig.~\ref{fig:schedules} left in orange shows over the training epochs via a cosine schedule. The right figure shows the effect of varying $\gamma$ on the $r$-span.

\section{Data and Implementation}
\label{sec:implementation}

\subsection{Training data}
\begin{figure}[tb]
    \centering
    \includegraphics[width=\columnwidth, clip, trim=80 110 70 120]{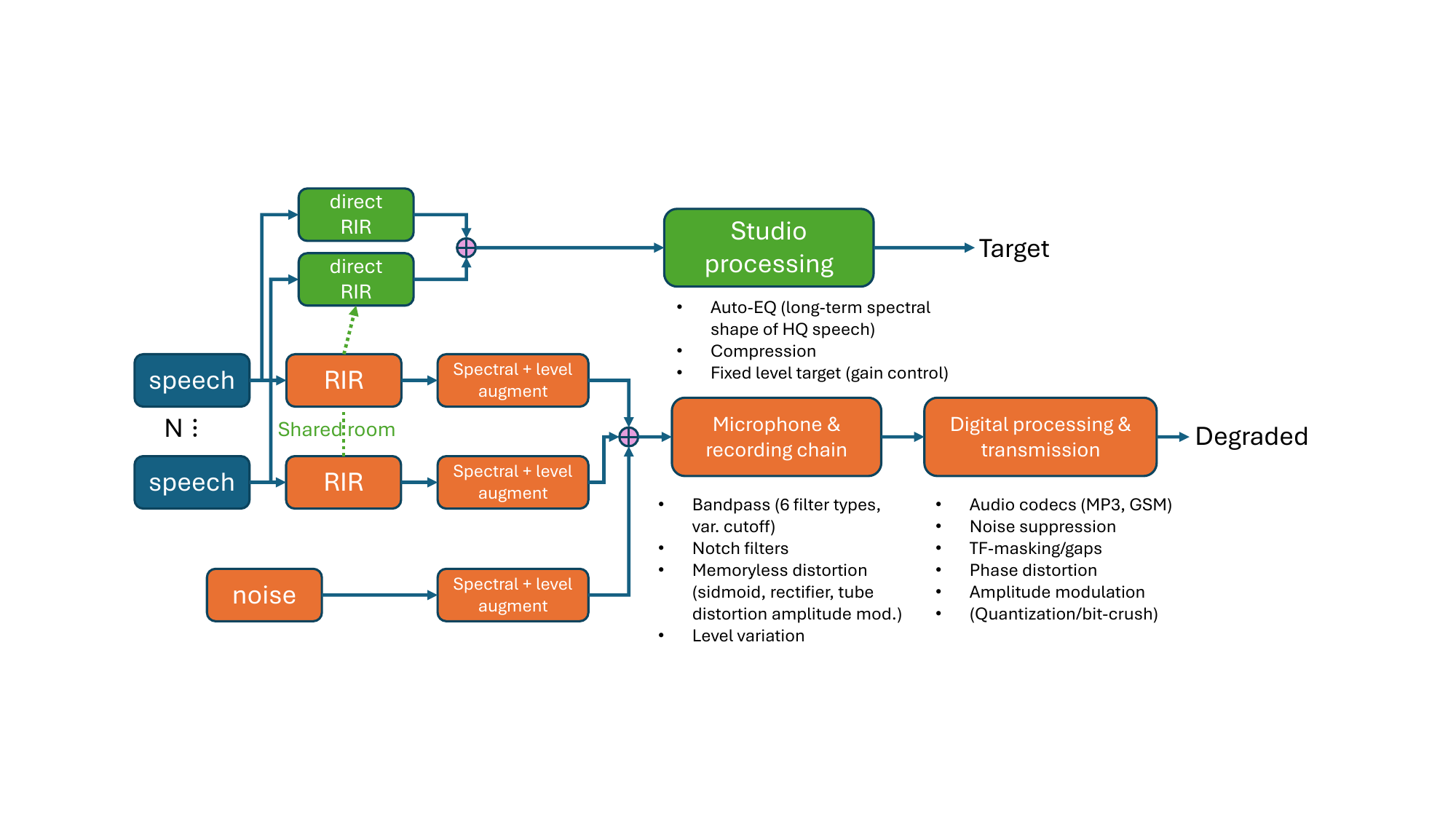}
    \vspace{-16pt}
    \caption{Data generation process for degraded input $y$ and target $x_0$}
    \label{fig:data_augmentation}
\end{figure}

We generate degraded and target speech pairs with on-the-fly augmentation with a similar pipeline as in \cite{Hsieh2026} using studio-quality clean speech from EARS \cite{Richter2024a}, reverberation via \acp{RIR} simulated by the image method \cite{Allen1979}, non-speech background noise from the DNS Challenge \cite{Reddy2021a}, and a wide variety of signal degradations shown in Fig~\ref{fig:data_augmentation}, outlined in the following: speech segments of 1--2 talkers are convolved with a \ac{RIR}, different \ac{RIR} per talker but within the same room, and added with random mixing gain. Background noise is added with a \ac{SNR} in $\mathcal{N}\{5, 10\}$~dB, and signal levels vary with $\mathcal{N}\{-40, 10\}$~dBFS. We apply simulated degradations to the noisy speech: bandwidth limitation with lower (100-800~Hz) and upper (1.5~kHz - Nyquist) cut-off frequency with 6 different filter types, notch filters, non-linear distortions (sigmoid, rectification, overdrive effect), audio codecs with variable bitrate (GSM, MP3), spectral bubble masking, phase distortion, amplitude modulation, over-aggressive stationary noise suppression, and audio dropouts in the range of 10-80~ms.
The speech target signals are obtained as time-aligned versions using direct-path only (anechoic) \ac{RIR} counterparts, and studio-like processing: target speech is normalized to -25~dBFS, auto-equalized to match the long-term spectral shape of studio-processed speech from DAPS \cite{Mysore2015}, and mild dynamic compression. 

\subsection{Network architectures}
We investigate three network architectures in this work, the first two are existing baselines, while the third is a novel design to balance latency, complexity and capaciy demands. All networks operate on the same complex compressed spectral representation \eqref{eq:complex_compressed_signal}, and concatenate signals $(\bm{x}_t, \bm{y})$ at the input channel dimension and accept a time embedding $t$ (and optionally $r$). 
Although \emph{NCSN++} \cite{Song2021} is the most widely used backbone for audio diffusion \ac{FM}, as pointed out in \cite{Hsieh2026,Welker2025}, the temporal downsampling at each encoder layer introduces a prohibitive algorithmic latency of about 600~ms. We term this architecture \emph{NCSN++ noncausal}, and include it only as upper bound baseline, as does not meet typical real-time requirements. As comparable fair baseline, we use the simple modification from \cite{Hsieh2026} by removing the temporal downsampling, termed \emph{NCSN++ causal}. Note that this incurs a significantly higher complexity due to larger internal feature maps from absent temporal downsampling. For fair comparison, we train NCSN++ models also with \ac{DP} loss \eqref{eq:data_prediction_loss}.

As second baseline, we use an updated version of the Convolutional 1D U-net with Gated Linear Units (\emph{ConvGLU1D}) from \cite{Hsieh2026}. We improved its performance by adding gradually increasing convolution dilation to increase the receptive field from originally 0.75~s to 2.1~s. While this model is reasonably fast and introduces no latency, we note that it limits the performance. 


\begin{table}[tb]
\centering
\vspace*{-10pt}
\caption{Model Comparison. MACs are for NFE=1.}
\label{tab:model_comparison}
\resizebox{\columnwidth}{!}{%
\begin{tabular}{@{}lrrrr@{}}
\toprule
\rowcolor[HTML]{FFFFFF} 
{\color[HTML]{000000} \textbf{model}} &
  \multicolumn{1}{l}{\cellcolor[HTML]{FFFFFF}{\color[HTML]{000000} \textbf{latency (ms)}}} &
  {\color[HTML]{000000} \textbf{params (M)}} &
  {\color[HTML]{000000} \textbf{MACs/s (G)}} &
  {\color[HTML]{000000} \textbf{context (s)}} \\ \midrule
\cellcolor[HTML]{D9D9D9}NCSN++ noncausal &
  \cellcolor[HTML]{F8696B}600 &
  \cellcolor[HTML]{FB9F76}53.0 &
  \cellcolor[HTML]{FFEB84}66.41 &
  \cellcolor[HTML]{D9D9D9}7.3 \\
NSCN++ causal \cite{Hsieh2026} &
  \cellcolor[HTML]{63BE7B}20 &
  \cellcolor[HTML]{FB9F76}53.0 &
  \cellcolor[HTML]{FFEA84}142.78 &
  0.61 \\
\cellcolor[HTML]{D9D9D9}DiffusionBuffer \cite{Lay2025} &
  \cellcolor[HTML]{FEC97E}180 &
  \cellcolor[HTML]{A7D17E}22.2 &
  \cellcolor[HTML]{F8696B}8810.00 &
  \cellcolor[HTML]{D9D9D9} \\
StreamFM   \cite{Welker2025} &
  \cellcolor[HTML]{FFEA84}32 &
  \cellcolor[HTML]{C3D980}27.9 &
  \cellcolor[HTML]{FFE884}282.00 &
   \\
\cellcolor[HTML]{D9D9D9}ConvGLU-1D \cite{Hsieh2026} &
  \cellcolor[HTML]{63BE7B}20 &
  \cellcolor[HTML]{F8696B}61.8 &
  \cellcolor[HTML]{63BE7B}0.10 &
  \cellcolor[HTML]{D9D9D9}2.11 \\
\textbf{RMFSR} &
  \cellcolor[HTML]{63BE7B}20 &
  \cellcolor[HTML]{63BE7B}7.8 &
  \cellcolor[HTML]{68BF7B}1.22 &
  2.13 \\ \bottomrule
\end{tabular}%
}
\end{table}
Therefore, we develop a new model to strike a balance between powerful enough learning capacity, but low enough complexity and no introduced latency. We use a 5 layer U-net using inverted residual bottleneck layers \cite{Sandler2018}, which inflate layer-internal features for depth-wise convolution 2x, and frequency attention. The encoder uses causal 3x3 convolutions (frequency x time) with increasing time dilation, the decoder 3x2 without dilation, and a 4-layer \ac{TCN} bottleneck with temporal only kernel (1x11) and increasing dilation. The encoder channel count is [64,64,128,256,256], mirrored at the decoder. Skip connections go from each encoder layer to its corresponding decoder via 1x1 convolution mapping and additive joining. The activation functions are SnakeBeta. The flow time embedding is implemented similar for all models: Gaussian Fourier embedding projected to 128 dimensions, added before each conv layer as conditioning. 
We term the resulting model \textbf{RMFSR} (Real-time Mean Flow Speech Restoration). Table~\ref{tab:model_comparison} compares the algorithmic latency (for the zero-latency architectures NCN++ causal, ConvGLU1D and RMFSR, the latency is only determined by the \ac{STFT} window size, here 20~ms), parameter count, \ac{MAC} per second of audio and temporal context a.k.a.\ receptive field.

\subsection{Flow Matching distribution design}
While a uniform distribution of the flow time $t$ is standard, a biased distribution towards the center and noisier steps often helps performance. For training, we sample using a logit-normal distribution with mean 0.4, i.\,e.\ $t \sim \operatorname{Sigmoid}(\mathcal{N}(0.4, 1))$.
Careful choice of the prior distribution of $\bm{x}_1$ can boost performance. As given in \eqref{eq:prior_x} we start from a Gaussian centered around $\bm{y}$. While the noise $\bm{\epsilon}$ is typically Gaussian (i.e. white with flat spectrum), we apply a $1/f$ energy decay to obtain pink noise, which ensures a more constant SNR over frequency, instead of burying high frequencies in noise. We use $\sigma_\text{max} = 0.3$ and $\sigma_\text{min}=10^{-8}$ for both white and pink noise.

\section{Experiments}
\label{sec:experiments}

\subsection{Test set and metrics}
We use the Signal Improvement Challenge 2024 (SIG2024) \cite{Ristea2025} test set, 500 real-world recordings with typical degradations from devices and VoiP processing. While most subjective MOS estimators are only moderately indicative for the rather novel speech restoration problem, we find DistillMOS \cite{Stahl2025} the most reliable, as it has been trained on a wide variety of tasks including speech generation. We additionally include the speech signal quality (SIG) predictor DNSMOS SIG \cite{Reddy2022}. 
As we find that MOS can stay rather unaffected by limited bandwidth, we add a an estimator of the \emph{spectral bandwidth} by estimating the maximum upper frequency $f_\text{max}$. This in combination with MOS gives a reliable indication if the model generates a full-band signal, but also at which quality. 
To assess hallucination, we compute \ac{WER} using Whisper Turbo\footnote{https://github.com/openai/whisper}.
Finally, we also conduct a crowd-sourced listening test on the best models using the SIG Challenge framework \cite{Ristea2025} using ITU P.804, rating the dimensions \emph{coloration, discontinuity, loudness, reverb, noise, signal quality and overall} with MOS.

\subsection{Results}
\begin{figure}[tb]
    \centering
    \includegraphics[width=\columnwidth, clip, trim=0 15 0 5]{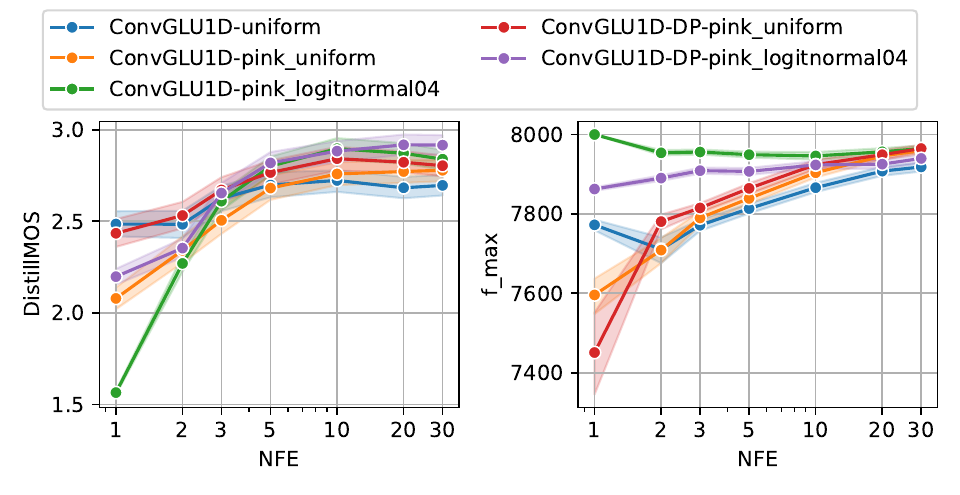}
    \vspace{-16pt}
    \caption{Impact of general FM improvements by time sampling (uniform vs.\ biased logit-normal), spectral noise distribution (white vs.\ pink) and data prediction (DP) loss.}
    \label{fig:ablation}
\end{figure}

We first show an ablation of the \ac{FM} design choices on the ConvGLU1D model. Figure~\ref{fig:ablation} shows model variants with standard velocity \ac{FM} loss \eqref{eq:fm_loss}, \ac{DP} loss \eqref{eq:data_prediction_conversion}, and flow time distribution uniform or skewed logit-normal, and pink vs.\ white noise $\bm{\epsilon}$.
We can see that pink noise modeling is better than white noise at higher NFE count, logit-normal mean 0.4 $t$ distribution is better than uniform $t$, and \ac{DP} loss outperforms the vanilla velocity loss. Note that the metrics at low NFE count become unreliable as different kinds of artifacts can behave irregular and even subjective ranking becomes debatable.

\begin{figure}[tb]
    \centering
    \includegraphics[width=\columnwidth, clip, trim=0 13 0 5]{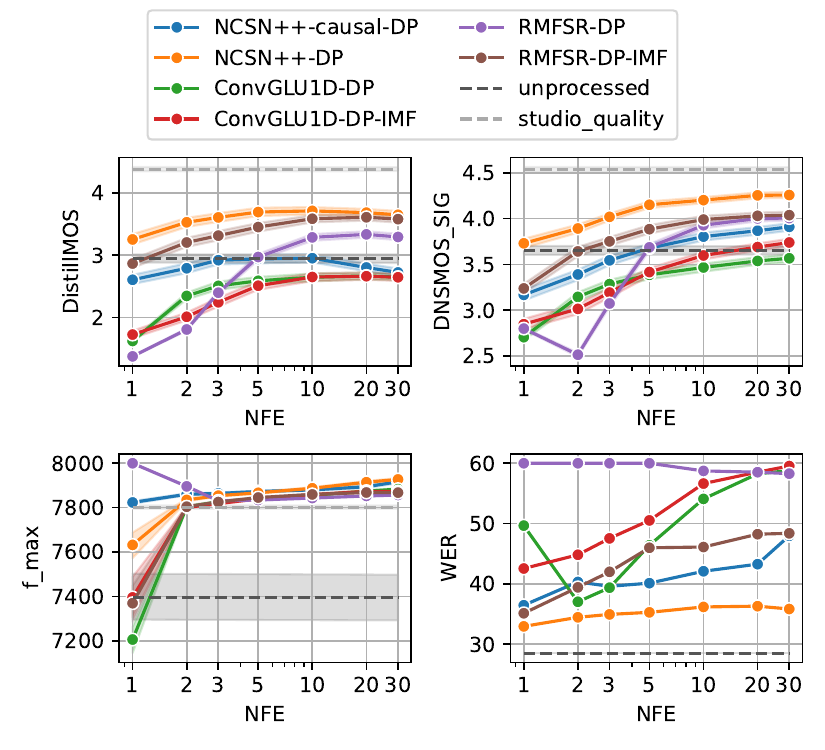}
    \vspace{-16pt}
    \caption{Objective results with baseline models.}
    \label{fig:results}
\end{figure}

The main results in Fig.~\ref{fig:results} show DistillMOS, DNSMOS SIG, the average maximum present frequency $f_\text{max}$, and WER over \ac{NFE}. References for unprocessed audio are shown as black-dashed, and a reference studio-quality speech dataset (DAPS produced) as grey-dashed lines.  
The first observation is that more powerful and complex models (see Tab.~\ref{tab:model_comparison}) generally perform better. \emph{NCSN++ noncausal} (orange), the large model with lookahead, performs best, while its causal counterpart (blue) drops in performance. Our proposed architecture \emph{RMFSR} trained with standard \ac{DP} \ac{FM} loss without Mean Flow (purple) outperforms NCSN++ causal (blue) at $NFE \!\!> \!\! 5$, but its performance significantly degrades at low NFE, where it also creates more noisy artifacts, increasing $f_\text{max}$, but highly degrading WER and MOS. However, when adding the \ac{DP}-\ac{IMF} training regime, the proposed model RMFSR-DP-IMF (brown) outperforms NCSN++ causal and comes very close to the noncausal top performer. There is still a gap in WER though. The weaker ConvGLU1D models underperform and struggle to provide significant MOS improvement over the input signal. While they are able to e.\,g.\, generate missing frequencies, balance the speech spectrum etc., their remaining artifacts degrade the audio quality. We further note that DP-\ac{IMF} loss does not show significant improvement for those weak models, although we observed improvement of IMF without \ac{DP} loss. Increasing \ac{WER} over NFE indicates more hallucination.

\begin{table}[tb]
\vspace{-10pt}
\caption{Subjective ITU P.804 listening test.}
\label{tab:p804}
\resizebox{\columnwidth}{!}{%
\begin{tabular}{@{}lccccccc@{}}
\toprule
\rowcolor[HTML]{FFFFFF} 
{\color[HTML]{000000} \textbf{model}} &
  {\color[HTML]{000000} \textbf{Coloration}} &
  {\color[HTML]{000000} \textbf{Discontinuity}} &
  {\color[HTML]{000000} \textbf{Loudness}} &
  {\color[HTML]{000000} \textbf{Noise}} &
  {\color[HTML]{000000} \textbf{Reverb}} &
  {\color[HTML]{000000} \textbf{Signal}} &
  {\color[HTML]{000000} \textbf{Overall}} \\ \midrule
\cellcolor[HTML]{D9D9D9}NCSN++-noncausal-DP &
  \cellcolor[HTML]{BEE3C9}3.68 &
  \cellcolor[HTML]{FBF2F5}3.96 &
  \cellcolor[HTML]{FBF4F7}3.96 &
  \cellcolor[HTML]{9FD7AF}4.05 &
  \cellcolor[HTML]{7BC890}4.36 &
  \cellcolor[HTML]{D5ECDD}3.59 &
  \cellcolor[HTML]{D3ECDC}3.20 \\
\textbf{RMFSR-DP-IMF} &
  \cellcolor[HTML]{D6EDDE}3.46 &
  \cellcolor[HTML]{FACACC}3.39 &
  \cellcolor[HTML]{A6D9B5}4.31 &
  \cellcolor[HTML]{77C78D}4.35 &
  \cellcolor[HTML]{73C589}4.41 &
  \cellcolor[HTML]{FBFCFE}3.28 &
  \cellcolor[HTML]{ECF6F1}2.91 \\
\rowcolor[HTML]{FCFCFF} 
\cellcolor[HTML]{D9D9D9}unprocessed &
  3.11 &
  4.10 &
  4.06 &
  3.34 &
  3.59 &
  3.27 &
  2.72 \\
NCSN++-causal-DP &
  \cellcolor[HTML]{FBE5E8}2.94 &
  \cellcolor[HTML]{F9B1B3}3.03 &
  \cellcolor[HTML]{FBEFF2}3.89 &
  \cellcolor[HTML]{FAD1D3}2.95 &
  \cellcolor[HTML]{B3DFC0}4.03 &
  \cellcolor[HTML]{FAB8BB}2.69 &
  \cellcolor[HTML]{F9A8AA}2.31 \\
\cellcolor[HTML]{D9D9D9}RMFSR-DP &
  \cellcolor[HTML]{FAC0C3}2.66 &
  \cellcolor[HTML]{F99D9F}2.75 &
  \cellcolor[HTML]{A9DBB7}4.30 &
  \cellcolor[HTML]{ABDBB9}3.96 &
  \cellcolor[HTML]{B8E1C4}4.00 &
  \cellcolor[HTML]{F9A5A7}2.52 &
  \cellcolor[HTML]{F98D90}2.18 \\ \bottomrule
\end{tabular}%
}
\vspace{-10pt}
\end{table}
Finally, we confirm the results by a subjective test using ITU P.804 as in the SIG Challenge 2024 \cite{Ristea2025} in Table~\ref{tab:p804}. Improvements are color coded relative to the unprocessed signals. The proposed model improves overall MOS by 0.2.
We note a general degradation in discontinuity, which could be attributed to occasional chopped off syllables in challenging conditions\footnote{Audio examples at \url{https://sebraun-msr.github.io/realtimemeanflowspeechrestoration/}}.
The proposed model outperforms the non-causal baseline in Noise, Reverb and Loudness, while there is some gap in other metrics, which is still a significant achievement given the 50x complexity and lookahead savings, and strongly outperforms the causal baseline on top of a 120x saving.

\section{Conclusions}
\label{sec:conclusions}
This work paved the way to drastically reduce computational cost and latency for general speech restoration flow-matching models. We demonstrate a 120x gain at increased quality by adopting Mean Flow training, careful designed flow-path trajectories and sampling, and a more cost efficient and latency-free convolutional architecture. We demonstrate the validity of results on a real-world public dataset, including subjective and objective audio quality metrics. Our resulting model performs close to the state-of-the-art non-causal model without its latency and complexity penalty. Despite these substantial gains, we find that under these constraints, one-step inference achieves still insufficient quality and it leaves room for improvement to close the gap on GANs and less for complicated deployment.

\balance
\bibliographystyle{IEEEbib}
\bibliography{O:/literature/sapref}

\end{document}